\documentclass[11pt,letterpaper,superscriptaddress]{JHEP3}
\usepackage{epsfig}
\usepackage{amssymb,amsmath,amsfonts,amstext,graphics}
\usepackage{slashed}
\usepackage{cite}
\newcommand{\beq}{\begin{eqnarray}}
\newcommand{\eeq}{\end{eqnarray}}

\def\ltap{\ \raise.3ex\hbox{$<$\kern-.75em\lower1ex\hbox{$\sim$}}\ }
\def\gtap{\ \raise.3ex\hbox{$>$\kern-.75em\lower1ex\hbox{$\sim$}}\ }
\newcommand{\gsim}{\lower.7ex\hbox{$\;\stackrel{\textstyle>}{\sim}\;$}}
\newcommand{\lsim}{\lower.7ex\hbox{$\;\stackrel{\textstyle<}{\sim}\;$}}

\def\be{\begin{equation}}
\def\ee{\end{equation}}
\def\bea{\begin{eqnarray}}
\def\eea{\end{eqnarray}}

\newcommand{\GeV}{\,\mathrm{GeV}}

\newcommand{\MU}{\ensuremath{M_{\mathcal{U}}}}
\newcommand{\duv}{\ensuremath{d_{UV}}}
\newcommand{\dU}{\ensuremath{d_{\mathcal{U}}}}
\newcommand{\OU}{\ensuremath{\mathcal{O}_{\mathcal{U}}}}

\newcommand{\Ouv}{\ensuremath{\mathcal{O}_{UV}}}
\newcommand{\Oir}{\ensuremath{\mathcal{O}_{IR}}}
\newcommand{\LambdaU}{\ensuremath{\Lambda_{\mathcal{U}}}}
\newcommand{\Lambdacft}{\ensuremath{\Lambda_{\slashed{\mathcal{U}}}}}

\def\unit{\relax{\rm 1\kern-.26em I}}

\newcommand{\Tr}{{\text{ Tr }}}

\title{Bounds on Unparticles from the Higgs Sector}
\author{Patrick J. Fox$^a$, Arvind Rajaraman$^b$ and Yuri Shirman$^b$\\
$^a$ Theoretical Physics Group, Lawrence Berkeley National
Laboratory, Berkeley, CA 94720
\\
$^b$ Department of Physics and Astronomy, University of
California, Irvine, CA 92697}
\preprint{LBNL-62680\\
UCI-TR-2007-23} \abstract{ We study supersymmetric QCD in the
conformal window as a laboratory for unparticle physics, and
analyze couplings between the unparticle sector and the Higgs
sector. These couplings can lead to the unparticle sector being
pushed away from its scale invariant fixed point.  We show that
this implies that low energy experiments will not be able to see
unparticle physics,
 and the best hope of seeing unparticles is in high energy
collider experiments such as the Tevatron and the LHC.  We  also
demonstrate how the breaking of scale invariance could be observed
at these experiments. }

\begin{document}

\section{Introduction and Conclusions}
Recently there has been a lot of interest
\cite{Liao:2007bx,Aliev:2007qw,Ding:2007bm,Chen:2007vv,Luo:2007bq,Cheung:2007ue,Li:2007by,Catterall:2007yx}
in unparticle theories \cite{Georgi:2007ek,Georgi:2007si} in which
the Standard Model (SM) is coupled to a conformal sector (called
the unparticle sector).  As shown in
\cite{Georgi:2007ek,Georgi:2007si}, the conformal sector can have
interesting and unexpected consequences. In this note, we shall
investigate the effects on the conformal sector from the Higgs
sector, and we will show that this leads to surprising new bounds
on unparticle physics.

The main point of our analysis is that the coupling to the Higgs
sector is the most important operator; in fact, if there is a
scalar unparticle operator of dimension less than 2, the coupling
to the Higgs sector is through a relevant operator. When the Higgs
gets a vacuum expectation value, this operator breaks the
conformal invariance of the hidden sector. For unparticle physics
to be relevant, this breaking scale is required to be sufficiently
low. (This operator may also lead to strong effects on Higgs
physics from unparticles, which would be interesting to
investigate.)

This requirement imposes strong constraints on the unparticle
sector. If these constraints are satisfied, low energy experiments
will not be able to probe any aspects of unparticle physics. The
only place where unparticle physics will be relevant is in high
energy experiments like the Tevatron and the LHC, which can indeed
probe unparticle physics. In fact, the breaking of conformal
invariance may also be measurable as deviations from the
predictions of unparticle physics.

We will begin by discussing a  model of unparticle physics, which
is different from the previously suggested models. We take the
conformal sector to be a supersymmetric gauge theory, which at low
energies flows to a conformal theory. Using supersymmetry, we will
be able to explicitly calculate the dimensions of chiral operators
in this theory, and show that this is a good model for unparticle
(and now also unsparticle) physics. We note that anomalous
dimensions in these theories can be large, in contrast to
Banks-Zaks fixed points which are weakly coupled.

We then couple this sector to the Standard Model, focusing on the
Higgs sector. We show that conformal invariance is broken at low
energies, and for reasonable choices of scales the inclusion of
the Higgs-unparticle operator means that low energy experiments
are unable to see the effects of unparticles.  Finally, we propose
a toy model for a theory with unparticles and a breaking of scale
invariance, and calculate experimental predictions of this effect.

\section{Supersymmetric QCD as a model of unparticle physics}

We would like to have an example of a conformal field theory in
which it is possible to do semi-quantitative calculations.  This
is challenging since we also require the theory be strongly
coupled in order that anomalous dimensions can be large.
Remarkably such an example exists in the literature
\cite{Seiberg:1994bz,Seiberg:1994pq}, in the form of
supersymmetric QCD (SQCD) in a certain regime.  We will therefore
consider SQCD in the conformal window as a laboratory for
unparticle physics.

We briefly review the results of SQCD (for a comprehensive review
see \cite{Intriligator:1995au}).  Consider SQCD with gauge group
$SU(N_C)$ and $N_F$ vector-like quark superfields ($Q$, $\bar{Q}$)
with $\frac{3}{2}N_C<N_F<3N_C$ (we call this the electric theory).
Such a theory flows to a strongly coupled conformal fixed point in
the infrared (IR).
 At the fixed point the theory
has a dual (magnetic) description, with gauge group $SU(N_F-N_C)$,
$N_F$ dual-quark superfields ($q$, $\bar{q}$), a gauge singlet
meson superfield $M$ (transforming in the bifundamental
representation of the $SU(N_F)\times SU(N_F)$ flavor symmetry, and
superpotential,
\begin{equation}
W_{mag}=\bar{q} M q \,.
\end{equation}
The meson of the magnetic description corresponds to the gauge
invariant composite $\bar Q Q$ of the electric theory.

The magnetic conformal theory can now be coupled to the Standard
Model, and will then be a candidate for the unparticle sector. In
general, we can write the ultraviolet (UV) coupling of an operator
of dimension $\duv$ in the unparticle sector to a SM operator of
dimension $l$ as
\begin{equation}\label{eqn:UVops}
 \frac{1}{\MU^{l+\duv-4}}\mathcal{O}_{SM}\Ouv
\end{equation} Below the strong coupling scale $\LambdaU$, these couplings flow to
\begin{equation}
\label{eqn:IRops} \mathcal{C}
\frac{\LambdaU^{\duv-\dU}}{\MU^{l+\duv-4}}\mathcal{O}_{SM}\Oir
\end{equation}
(in the notation of Georgi \cite{Georgi:2007ek,Georgi:2007si}
$k=l+\duv-4$).

Supersymmetric QCD allows us to make this explicit. For example,
adding a superpotential coupling in the UV regime of the magnetic
description
\begin{equation}
 W_c= \frac{1}{\MU}H L \bar e \Tr M
\end{equation}
leads, among others, to the following terms in in the Lagrangian
\begin{equation}
 \frac{1}{\MU}H L \bar e \Tr M + \left(\frac{1}{\MU} H \tilde L \tilde{\bar e} q^*
\bar q^* + h.c.\right)\,,
\end{equation}
which have the form (\ref{eqn:UVops}) with $\Ouv= \Tr M$ of
dimension $\duv=1$ and $\Ouv= q\bar q$ of dimension $\duv=2$.
Below the strong coupling scale, once the theory reaches its
conformal fixed point, the dimensions of these operators can be
computed from their R-charges to be $\dU=3\frac{N_C-NF}{N_F}$ and
$\dU=3\frac{N_C}{N_F}$ respectively.  In the conformal window, the
dimension of both operators lie between 1 and 2, making them
perfect candidates for the operator $\OU$ of the unparticle
conformal sector \cite {Georgi:2007ek}.

A couple of comments are in order:
\begin{enumerate}
\item For generic choices of $N_C$ and $N_F$, in the conformal
window, the dimensions of the operators $q\bar q, M$ significantly
differ from integer values. This is unlike the Banks-Zaks (BZ)
theory \cite{Banks:1981nn} which has a weakly coupled fixed point
where all operators have dimension close to their classical value.
In particular, all gauge invariant operators in BZ theory have
almost integer dimensions. \item We can perturb the theory by
adding a term to the action $\lambda\Oir=\lambda\Tr M$. This
corresponds to adding a mass term for the quarks of the electric
description. The result of this mass term is that at low energies,
the quarks can be integrated out, and the theory becomes a pure
super-Yang Mills theory, which is no longer conformal.
\end{enumerate}

\section{Operator analysis and experimental constraints}

The couplings of the unparticle sector and the SM sector can have
interesting effects. Most interest thus far has concentrated on
operators involving SM fermions and gauge bosons (with the goal of
determining low energy signatures of unparticles) and consequently
on operators with $l\ge 3$ . Because the operator of lowest
dimension in the unparticle sector has dimension greater than $1$,
this means that the coupling operator is irrelevant ($l+\duv-4\ge
0$).

However there is another type of coupling between the SM and the
unparticle sector, involving the SM Higgs boson. The coupling is
of the form
\begin{equation}
 \frac{1}{\MU^{\duv-2}}|H|^2 \Ouv
\end{equation}
which flows in the infrared (IR) to
\begin{equation}\label{eqn:higgsupsop}
 C_{\mathcal{U}}\frac{\LambdaU^{\duv-\dU}}{\MU^{\duv-2}}|H|^2 \Oir
\end{equation}
In the following we will assume that $\Ouv$ and $\Oir$ are the same operators as in 
eqns. (\ref{eqn:UVops}) and (\ref{eqn:IRops}) respectively. 
The dimension of $\Oir$ is usually assumed to lie between 1 and 2
\cite{Georgi:2007si}, as is indeed the case for SQCD. For such
operators, this coupling is {\it relevant} in the CFT and can
significantly change the low energy physics of the unparticle
sector.  We note there is no symmetry that can forbid this
operator without simultaneously forbidding fermion and gauge boson
operators coupling to the unparticle sector.

We note that if there is no scalar operator of dimension less than
2 in the unparticle sector, then the operator
(\ref{eqn:higgsupsop}) is irrelevant.  Any operator with dimension
less than 2 would then have to be a vector or higher tensor
operator. Such scenarios are difficult to realize in SQCD, but may
be realized in more exotic theories. (For example, one may use
AdS/CFT and consider the CFT dual of an AdS theory which only
contains vector fields.) We will not consider this possibility
further.

Once the Higgs acquires a vev, the operator (\ref{eqn:higgsupsop})
introduces a scale into the CFT.  This relevant operator will
cause the unparticle sector to flow away from its conformal fixed
point and the theory will become non-conformal at a scale
\Lambdacft, where
\begin{equation}
\label{eqn:lowestlambda} \Lambdacft^{4-\dU} =
\left(\frac{\LambdaU}{\MU}\right)^{\duv-\dU} \MU^{2-\dU} v^2 \,.
\end{equation}
Below this scale the unparticle sector presumably becomes a
traditional particle sector.  For consistency we require
$\Lambdacft < \LambdaU$. If there is to be any sense in which the
theory is truly conformal the two scales should be well separated.

Breaking of the conformal invariance due to the new operators has
important implications for unparticle phenomenology. For any given
experiment, unparticle physics will only be relevant if
\begin{equation}
\label{eqn:Qconstraint}
 \Lambdacft<Q
\end{equation}
where $Q$ is the typical energy of the experiment. For lower
energies, the unparticle sector can be treated as a particle
sector.  The constraint of (\ref{eqn:Qconstraint}) then takes the
form
\begin{equation}
Q^{4-\dU} > \left(\frac{\LambdaU}{\MU}\right)^{\duv-\dU}
\MU^{2-\dU} v^2 \,.
\end{equation}

This suggests that low energy experiments may not be sensitive to
unparticle physics.  To see this explicitly we note that any
observable effect of the operator (\ref{eqn:IRops}) will be
proportional to
\begin{equation}
\label{eqn:Obound}
\epsilon=\left(\frac{\LambdaU}{\MU}\right)^{2\duv-2\dU}\left(\frac{Q}{\MU}\right)^{2(\dU+l-4)}
\,.
\end{equation}
Then the effects of the unparticle sector on observables are
bounded by
\begin{equation}
\label{eqn:expconstraints}
\epsilon<\left(\frac{Q}{\MU}\right)^{2l}\left(\frac{\MU}{v}\right)^4
\,.
\end{equation}
It is interesting that this constraint is completely independent\footnote{This
is not entirely true as there is dependence on $\dU$ due to the
modification of phase space as well as dimensionless couplings in the Lagrangian.  
This results in corrections of order
1, but the dependence on energy scales remains the same.} of both
the UV and IR scaling dimension of the CFT operator and the
potential effects of the unparticle sector are constrained by only
3 parameters: the experimental energy, scale of the electroweak
symmetry breaking and the energy scale at which the interactions
between the SM and unparticle sector are generated.

Let us now concentrate on effects  on (g-2), the anomalous
magnetic moment of the electron. In this case the SM operator is
simply\footnote{There is also the possibility of a pseudoscalar
operator $\bar{e}\gamma_5 e$ but this follows the same scaling
arguments.} $\mathcal{O}_{SM}=\bar e e $ and the relevant energy
scale is $m_e$. Therefore eqn. (\ref{eqn:expconstraints}) becomes
\begin{equation}
 \epsilon<\frac{m_e^6}{\MU^2v^4}\,.
\end{equation}
Since effective field theory works at the electroweak scale, we
expect $\MU\gtap 100\GeV$. We then find $\epsilon < 10^{-28}$.
This should be contrasted with the existing experimental
constraint given in \cite{Gabrielse:2006gg}, $\epsilon <
10^{-11}$. Therefore the effects of the unparticle sector are
completely invisible in (g-2) experiments.

It is clear from (\ref{eqn:Obound}) that signals of unparticle
physics increase with energy and the LHC is the most promising
place where unparticle physics can be discovered. For such an
experimental discovery to be possible, $\MU$ should not be too
high. Assuming that we can detect deviations from the Standard
Model of order $\epsilon\sim 1\%$, we see that unparticle physics
will be visible at the LHC as long as $\MU\simeq 10^5 \GeV$.

\begin{figure}[t]
\begin{center}
\includegraphics[width=.45\columnwidth]{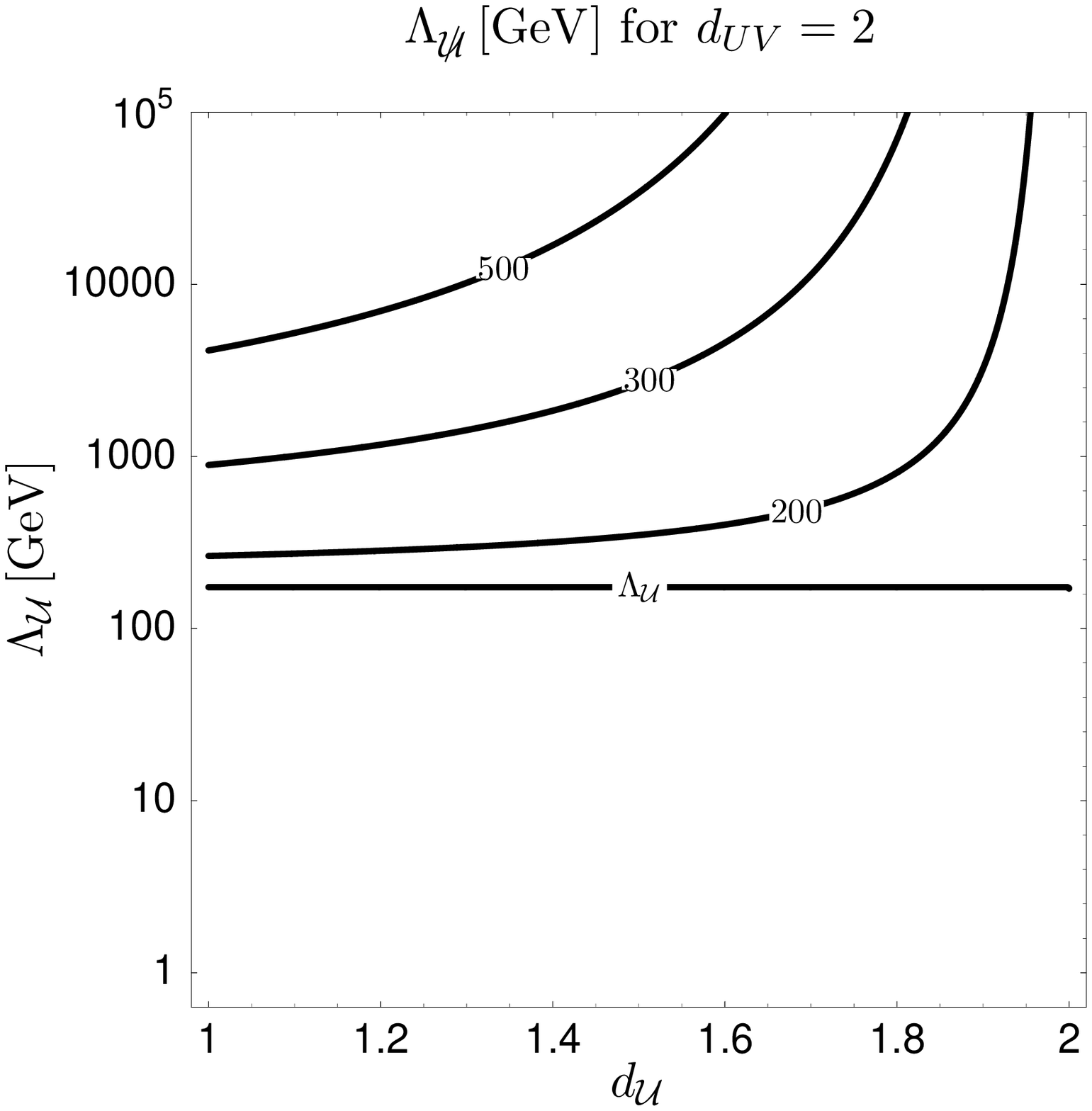}
\hspace{0.05\columnwidth}
\includegraphics[width=.45\columnwidth]{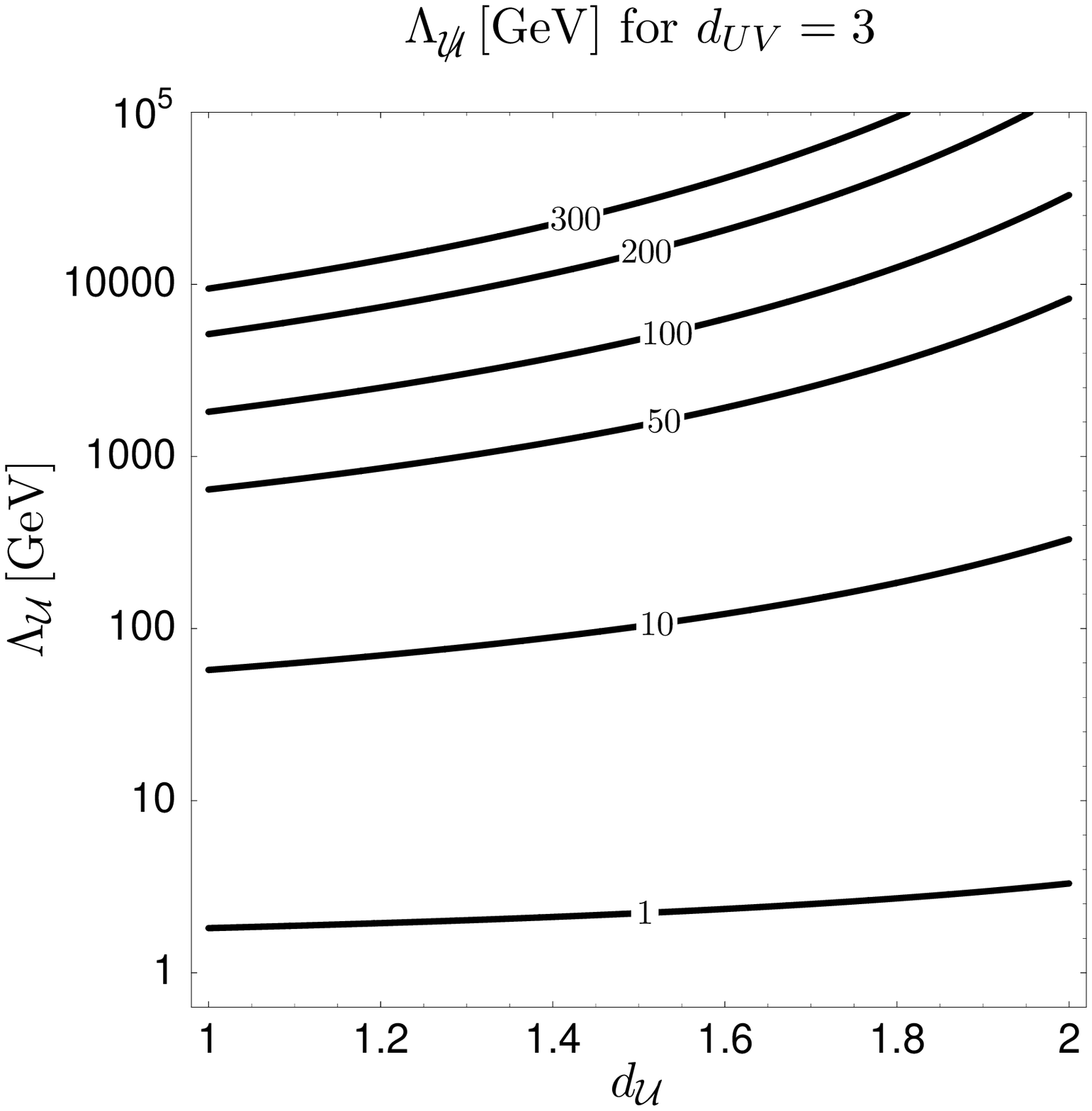}
\caption{Contours of fixed CFT breaking scale, $\Lambdacft$, as a
function of the IR dimension of the unparticle operator, $\dU$,
and the scale at which it becomes conformal, $\LambdaU$.   Two
particular choices for the UV dimension of the unparticle operator
are made, $\duv=2,3$. As discussed in the text we only consider
$\Lambdacft\le \LambdaU$.}\label{fig:lambdacftplots}
\end{center}
\end{figure}

If SQCD is the model of the unparticle sector $\duv$ is either 1
or 2. Using (\ref{eqn:lowestlambda}) we can easily see that the
case of $\duv=1$ typically predicts breaking scales well above
energies probed at the LHC making unparticles irrelevant even
there.  However, this particular coupling may not be present or
there may exist other realizations of unparticles in which
$\duv=1$ is not allowed, in which case it is still possible to be
at the fixed point for LHC energies.  In
Figure~\ref{fig:lambdacftplots} we show how the breaking scale
$\Lambdacft$ varies with $\dU$ and $\LambdaU$ for two different
choices of $\duv$; in both cases $\MU=10^5\,\GeV$, although for
$\duv=2$ there is no dependence on $\MU$.  From these plots we can
see that the range over which the unparticle sector is conformal
can be made large only at the expense of increasing the UV
dimension of the unparticle operator or the scale $\MU$.  The
former then requires very large anomalous dimensions to allow
$\dU$ to lie between 1 and 2, and the latter may make the
discovery of unparticles beyond the reach of any experiment. On
the other hand, as we will see in the next section if $\Lambdacft$
is not very small, then the deviations from both particle and
unparticle physics may be measurable.

\section{New effects in non-unparticle physics}

In a pure conformal theory, the correlator can be written as \be
\langle
 O_{\cal U}(x)O_{\cal U}(0)\rangle =\int \frac{d^4P}{(2\pi)^4}
e^{-ipx}|\langle0|O_{\cal U}|P\rangle|^2 \rho(P^2) \ee with \be
|\langle0|O_{\cal U}|P\rangle|^2 \rho(P^2) =A_{d_{\cal
U}}\theta(P^0)\theta(P^2)(P^2)^{d_{\cal U}-2}\,. \ee

We propose a simple toy  model where conformal invariance is
broken at a low energy $\mu$ by modifying the above equation to
\be |\langle0|O_{\cal U}|P\rangle|^2 \rho(P^2) =A_{d_{\cal
U}}\theta(P^0)\theta(P^2-\mu^2)(P^2-\mu^2)^{d_{\cal U}-2}\,. \ee
This modification corresponds to shifting the spectrum to remove
modes with energy less than $\mu$.

This model maintains the unparticle nature of the hidden sector
while including the effects of the breaking of scale invariance.
While it is not clear whether  such a modification can arise from
a consistent QFT,  this represents a simple model to study effects
of deviations from conformal invariance. More generally, once
scale invariance is broken, there may be particle-like modes that
would appear as isolated poles in the spectral function; we ignore
these effects in this simple model.  It would be interesting to
see if consistent models of this type can be constructed and what
features they possess\footnote{We thank C. Csaki and H. Georgi for
emphasizing this point to us.}.

\begin{figure}[t]
\begin{center}
\includegraphics[width=0.7\columnwidth]{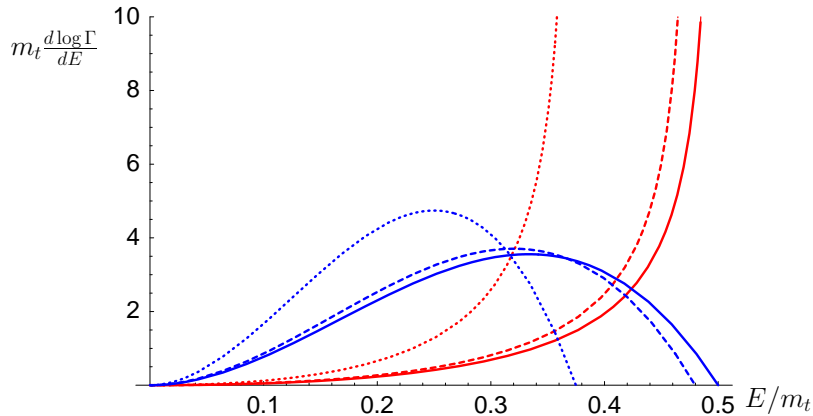}
\caption{The differential decay rate, $m_t\frac{d\log
\Gamma}{dE_u}$, for the decay $t\rightarrow u\,\OU$ as a function
of final state quark energy.  The red curves (concave) assume the
IR dimension of the unparticle operator is $\dU=4/3$ and the blue
(convex) assume $\dU=3$.  In both cases the solid, dashed and
dot-dashed curves label the scale, $\mu$, below which there are no
unparticle modes and correspond to $\mu/m_t=0,0.2,0.5$
respectively.}\label{fig:shape}
\end{center}
\end{figure}

This modification can produce observable effects. To illustrate
this, we will reconsider the effects of unparticle physics on the
decay of the top through processes like $t\rightarrow u\,\OU$. The
decay rate for this process can be computed following
\cite{Georgi:2007ek} to be \be m_t\frac{d\log
\Gamma}{dE_u}=4\dU(\dU^2-1)\left(\frac{m_t}{M}\right)^6
\left(\frac{E_u}{m_t}\right)^2
\left(1-2\left(\frac{m_t}{M}\right)^2
\frac{E_u}{m_t}\right)^{\dU-2} \ee with $M^2=m_t^2-\mu^2$.  In
Figure~\ref{fig:shape} we show this modification for various
choices of $\mu$ and $\dU$.  Notice that the end point of the
distribution is no longer $m_t/2$ but is now
$\frac{m_t^2-\mu^2}{2m_t}$ and that the normalization of the
distribution changes.

It may also be possible to look for effects of the breaking of
scale invariance in other collider signatures, for instance the
interference between unparticle and SM propagators in simple
processes such as $e^+ e^-\rightarrow \mu+ \mu^-$
\cite{Georgi:2007si}.  It would  be especially interesting to see
how the inclusion of the coupling between the SM Higgs and the
unparticle sector affects Higgs physics. These questions are left for future work.

\section*{Acknowledgements}

We thank Howard Georgi for comments on the manuscript.  We thank
M. Bander for discussions.  The work of AR is supported in part by
NSF Grant PHY--0354993.  PJF was supported in part by the
Director, Office of Science, Office of High Energy and Nuclear
Physics, Division of High Energy Physics, of the US Department of
Energy under contract DE-AC02-05CH11231. PJF would like to thank
the Particle Theory group at UC Irvine for hospitality while part
of this work was completed.

\bibliography{unparticles}

\providecommand{\href}[2]{#2}\begingroup\raggedright\begin{thebibliography}{10}

\bibitem{Liao:2007bx}
Y.~Liao, {\it Bounds on unparticles couplings to electrons: From electron g-2
  to positronium decays},  \href{http://xxx.lanl.gov/abs/0705.0837}{{\tt
  arXiv:0705.0837 [hep-ph]}}.

\bibitem{Aliev:2007qw}
T.~M. Aliev, A.~S. Cornell, and N.~Gaur, {\it Lepton flavour violation in
  unparticle physics},  \href{http://xxx.lanl.gov/abs/0705.1326}{{\tt
  arXiv:0705.1326 [hep-ph]}}.

\bibitem{Ding:2007bm}
G.-J. Ding and M.-L. Yan, {\it Unparticle physics in dis},
  \href{http://xxx.lanl.gov/abs/0705.0794}{{\tt arXiv:0705.0794 [hep-ph]}}.

\bibitem{Chen:2007vv}
C.-H. Chen and C.-Q. Geng, {\it Unparticle physics on cp violation},
  \href{http://xxx.lanl.gov/abs/0705.0689}{{\tt arXiv:0705.0689 [hep-ph]}}.

\bibitem{Luo:2007bq}
M.~Luo and G.~Zhu, {\it Some phenomenologies of unparticle physics},
  \href{http://xxx.lanl.gov/abs/0704.3532}{{\tt arXiv:0704.3532 [hep-ph]}}.

\bibitem{Cheung:2007ue}
K.~Cheung, W.-Y. Keung, and T.-C. Yuan, {\it Novel signals in unparticle
  physics},  \href{http://xxx.lanl.gov/abs/0704.2588}{{\tt arXiv:0704.2588
  [hep-ph]}}.

\bibitem{Li:2007by}
X.-Q. Li and Z.-T. Wei, {\it Unparticle physics effects on d-anti-d mixing},
  \href{http://xxx.lanl.gov/abs/0705.1821}{{\tt arXiv:0705.1821 [hep-ph]}}.

\bibitem{Catterall:2007yx}
S.~Catterall and F.~Sannino, {\it Minimal walking on the lattice},
  \href{http://xxx.lanl.gov/abs/0705.1664}{{\tt arXiv:0705.1664 [hep-lat]}}.

\bibitem{Georgi:2007ek}
H.~Georgi, {\it Unparticle physics},
  \href{http://xxx.lanl.gov/abs/hep-ph/0703260}{{\tt hep-ph/0703260}}.

\bibitem{Georgi:2007si}
H.~Georgi, {\it Another odd thing about unparticle physics},
  \href{http://xxx.lanl.gov/abs/0704.2457}{{\tt arXiv:0704.2457 [hep-ph]}}.

\bibitem{Seiberg:1994bz}
N.~Seiberg, {\it Exact results on the space of vacua of four-dimensional susy
  gauge theories},  {\em Phys. Rev.} {\bf D49} (1994) 6857--6863,
  [\href{http://xxx.lanl.gov/abs/hep-th/9402044}{{\tt hep-th/9402044}}].

\bibitem{Seiberg:1994pq}
N.~Seiberg, {\it Electric - magnetic duality in supersymmetric nonabelian gauge
  theories},  {\em Nucl. Phys.} {\bf B435} (1995) 129--146,
  [\href{http://xxx.lanl.gov/abs/hep-th/9411149}{{\tt hep-th/9411149}}].

\bibitem{Intriligator:1995au}
K.~A. Intriligator and N.~Seiberg, {\it Lectures on supersymmetric gauge
  theories and electric- magnetic duality},  {\em Nucl. Phys. Proc. Suppl.}
  {\bf 45BC} (1996) 1--28, [\href{http://xxx.lanl.gov/abs/hep-th/9509066}{{\tt
  hep-th/9509066}}].

\bibitem{Banks:1981nn}
T.~Banks and A.~Zaks, {\it On the phase structure of vector-like gauge theories
  with massless fermions},  {\em Nucl. Phys.} {\bf B196} (1982) 189.

\bibitem{Gabrielse:2006gg}
G.~Gabrielse, D.~Hanneke, T.~Kinoshita, M.~Nio, and B.~Odom, {\it New
  determination of the fine structure constant from the electron g value and
  qed},  {\em Phys. Rev. Lett.} {\bf 97} (2006) 030802.

\end{thebibliography}\endgroup
\bibliographystyle{JHEP}

\end{document}